\documentclass[journal=ancac3,manuscript=article]{achemso}

\usepackage[version=3]{mhchem} 
\usepackage[T1]{fontenc} 
\usepackage[hidelinks]{hyperref}
\usepackage[mathletters]{ucs}
\usepackage[utf8]{inputenc}

\newcommand{\sectionsize}{\normalsize}

\author{Haonan Zong}
\affiliation{Department of Electrical and Computer Engineering, Boston University, Boston, MA 02215, USA}
\altaffiliation{Contributed equally to this work.}

\author{Celalettin Yurdakul}
\affiliation{Department of Electrical and Computer Engineering, Boston University, Boston, MA 02215, USA}
\altaffiliation{Contributed equally to this work.}

\author{Yeran Bai}
\affiliation{Department of Electrical and Computer Engineering, Boston University, Boston, MA 02215, USA}

\author{Meng Zhang}
\affiliation{Department of Biomedical Engineering, Boston University, Boston, MA 02215, USA}

\author{M. Selim \"Unl\"u}
\affiliation{Department of Electrical and Computer Engineering, Boston University, Boston, MA 02215, USA}
\alsoaffiliation{Department of Biomedical Engineering, Boston University, Boston, MA 02215, USA}
\email{selim@bu.edu}

\author{Ji-Xin Cheng}
\affiliation{Department of Electrical and Computer Engineering, Boston University, Boston, MA 02215, USA}
\alsoaffiliation{Department of Biomedical Engineering, Boston University, Boston, MA 02215, USA}
\email{jxcheng@bu.edu}

\title{Background-Suppressed High-Throughput Mid-Infrared Photothermal Microscopy via Pupil Engineering}

\begin{document}



\begin{abstract}
    \normalsize
    Mid-infrared photothermal (MIP) microscopy has been a promising label-free chemical imaging technique for functional characterization of specimens owing to its enhanced spatial resolution and high specificity. Recently developed wide-field MIP imaging modalities have drastically improved speed and enabled high-throughput imaging of micron-scale subjects. However, the weakly scattered signal from sub-wavelength particles becomes indistinguishable from the shot-noise as a consequence of the strong background light, leading to limited sensitivity. Here, we demonstrate background-suppressed chemical fingerprinting at a single nanoparticle level by selectively attenuating the reflected light through pupil engineering in the collection path. Our technique provides over three orders of magnitude background suppression by quasi-darkfield illumination in epi-configuration without sacrificing lateral resolution. We demonstrate 6-fold signal-to-background noise ratio improvement, allowing for simultaneous detection and discrimination of hundreds of nanoparticles across a field of view of 70 µm x 70 µm. A comprehensive theoretical framework for photothermal image formation is provided and experimentally validated with 300 and 500~nm PMMA beads. The versatility and utility of our technique are demonstrated via hyperspectral dark-field MIP imaging of \textit{S. aureus} and \textit{E. coli} bacteria.
\end{abstract}
\section{\sectionsize KEYWORDS} 
photothermal imaging, chemical imaging, nanoparticle detection, biosensing, bacteria detection, infrared spectroscopy \par

\section{\sectionsize INTRODUCTION} 

Vibrational spectroscopic imaging has been an essential tool for molecular fingerprinting in biological and medical sciences~\cite{ChengScienceReview}. Intrinsic molecular vibrations in specimens can be utilized as contrast by either optical absorption or inelastic Raman scattering. Therefore, the vibrational signal avoids challenges associated with fluorescence labels such as sample perturbation, photo-bleaching, and phototoxicity. The spontaneous Raman scattering spectromicroscopy offers sub-micron spatial resolution but the non-linear process limits sensitivity and acquisition speed~\cite{RamanScattering}. To boost the imaging speed, coherent Raman scattering microscopy is developed~\cite{SRS, CoherentRaman}. Historically, Fourier transform infrared (FTIR) microscopy has been one of the most prevalent chemical analysis tools in various fields ~\cite{FTIR1,FTIR2}. However, FTIR has a low spatial resolution of several microns due to the diffraction limit of the long illumination wavelengths in the mid-IR region (2.5 - 25 µm). Such resolution is insufficient to study intra-cellular structures inside a biological cell. Atomic force microscope infrared spectroscopy (AFM-IR) techniques beat this inherent resolution limit~\cite{AFMIR1,AFMIR2} by detecting thermal expansion with an AFM tip and nanoscale (20~nm) resolution has been achieved~\cite{AFMIR3}. Yet, the slow scanning process in AFM-IR limits imaging throughput. Additionally, applying AFM-IR to a liquid sample is very difficult, which limits its applications to characterize live cells. To tackle these challenges, mid-infrared photothermal (MIP) microscopy was recently developed, where vibrational absorption-induced chemical contrast was detected by a visible probe beam~\cite{Delong2016, Zhongming2017, epi2017}. After the initial inception in 2016~\cite{Delong2016}, MIP has been progressively evolving from the proof of concepts to commercialization and wide applications~\cite{miragePharma,mirageCrystals, mirageNeuron, mirageAtmospheric}. Technically, Delong Zhang \textit{et al.} first demonstrated sub-micron resolution mid-IR spectroscopic imaging of living cells and organisms with a point scanning approach in transmission mode~\cite{Delong2016}. An epi-detection MIP was subsequently developed to extend the system capability to opaque samples including drug tablets~\cite{epi2017}. The photothermal imaging in counter propagation detection geometry was also developed to employ high-NA objectives for high-resolution ($\sim$ 300~nm) and sensitive chemical detection down to 100~nm polystyrene beads at the high-temperature rise (30 to 45 K)~\cite{Zhongming2017, Masaru1ContrastMechanism, Masaru3Review} and demonstrated in material science application\cite{Masaru2Material,Masaru4Solar}. More recently, confocal Raman spectroscopy is integrated into MIP~\cite{IRaman}, providing complementary chemical information by simultaneous Raman and IR spectroscopy of specimens at co-registered sub-micron spatial resolution. Alternatively, mid-IR vibrational signatures have been extracted using a near-IR~\cite{Sander2} or shortwave IR laser probe~\cite{PanagisOE}.

Despite its high sensitivity, the scanning MIP microscopy approaches have three main challenges: (i) Limited imaging speed due to the pixel-by-pixel acquisition; (ii) Wavelength dependent focused spot size mismatch between the IR and visible beams; (iii) Mechanical instabilities and sample drift. Harnessing complementary metal-oxide-semiconductor (CMOS) cameras that provide ultra-sensitive (\textgreater 70~dB) and high-speed image (100s FPS) capturing at very low-cost, very recently introduced wide-field MIP systems~\cite{Yeran2019,phase2019,japanPhaseContrastSR,japanQPIOL,japanQPIOptica,Rohit_PNAS_2020} enabled high-throughput chemical imaging with a significant speed improvement. Bai \textit{et al.}~\cite{Yeran2019} developed a virtual lock-in camera technique to detect the photothermal induced interferometric reflectance change of sample placed on a silicon substrate. Similarly, Schnell \textit{et al.}~\cite{Rohit_PNAS_2020} reported an interferometric method for measuring surface thermal expansion on histopathology slides using a special common-path interferometry objective, namely, Mirau objective along with an extremely high full-well capacity camera (2 million e$^-$). Moreover, quantitative chemical phase microscopy of living cells has been demonstrated by reconstructing a very minute optical phase path length difference~\cite{phase2019, japanQPIOL, japanQPIOptica}. Despite these recent advancements, wide-field MIP imaging of single sub-wavelength particles such as bacteria and viruses has not been demonstrated. A major limitation is that these techniques rely on bright-field sample illumination in which weakly scattered light from sub-wavelength (\textless 500~nm) structures are overwhelmed by the large background arising from the strong illumination field. As the particle size decreases, the photothermal signal becomes buried under the background shot-noise from this unperturbed illumination. Thus, chemical imaging of nanoparticles becomes very difficult to achieve. 

Here, we reported a background-suppressed wide-field MIP microscope via dark-field geometry in the epi-illumination configuration without sacrificing resolution. Background suppression via dark-field geometry has been proved to drastically improves the signal-to-background ratio such that the camera captures only the scattered light~\cite{darkfield}. The dark-field illumination in the epi-configuration is achieved by selectively blocking the back-reflected light in the collection arm after the objective. Our approach obviates the need for special objectives or condensers as in the off-the-shelf dark-field microscopes~\cite{MertzBook}. The dark-field illumination through the objective can be performed via oblique illumination which is blocked using a field stop~\cite{objectiveTypeOblique1,objectiveTypeOblique2} and on-axis illumination which is blocked by a rod mirror~\cite{objectiveTypeLowNA1} or a circular stop~\cite{objectiveTypeLowNA2}. The objective-type on-axis dark-field illumination blocks only the small fraction of the objective numerical aperture (NA), low-NA part centered around the optical axis. Such illumination can employ high-NA objectives compared to that of the oblique, allowing for more sensitive and high-resolution detection of the back-scattered light from small specimens down to single fluorescent molecules~\cite{objectiveTypeLowNA2}. In this study, we illuminate the sample with a nearly collimated beam which is refocused at the objective back pupil after the specular reflection from the substrate surface. A custom fabricated blocker filters out this beam at the pupil's conjugate plane, thus, nearly no reflected light reaches the camera. We demonstrate more than 6-fold signal-to-background noise ratio improvement over a large field of view (FOV) of 70 µm $\times$ 70 µm, enabling simultaneous photothermal imaging of hundreds of particles at once. We establish a complete physical model for the photothermal image formation that utilizes boundary element methods and angular spectrum representation framework. The theory is experimentally validated with 300 and 500~nm PMMA beads. We further provide the transient temperature response for these beads by employing the time-gated pump-probe approach~\cite{phase2019}. To highlight its potential for biological applications, we demonstrate fingerprinting of single \textit{E. coli} and \textit{S. aureus} in the wide-field sense. Our method advances high-throughput nanoparticle imaging and characterization with chemical specificity at a single particle level.

\section{\sectionsize RESULTS AND DISCUSSION}
\subsection{\sectionsize Photothermal contrast mechanism in dark-field MIP}

\begin{figure}[!ht]
	\begin{center}
	\includegraphics[width=0.5\textwidth]{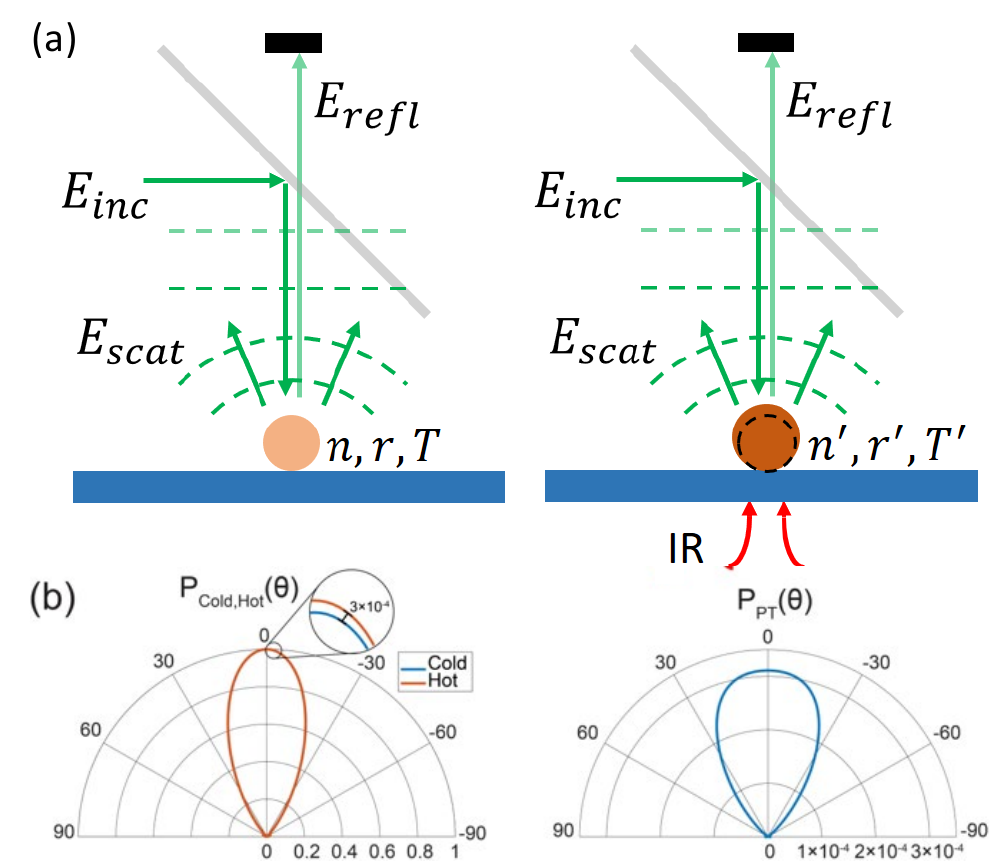}
	\caption{\textbf{Photothermal signal formation in dark-field detection.} (a) Photothermal contrast and detection mechanism for particles on top a substrate. (b) (Left) Simulated scattering polar plots of 500~nm PMMA bead on silicon substrate in the hot $\mathrm{P_{Hot}({\theta})}$ and cold $\mathrm{P_{cold}({\theta})}$ states and (Right) corresponding photothermal polar plot obtained by subtracting hot and cold states, $\mathrm{P_{PT}({\theta}) = P_{Hot}({\theta}) - P_{Cold}({\theta})}$. Signals are normalized by the maximum intensity value in the cold state. Polar plots are calculated from the far-field scattered fields obtained from the BEM simulations. The simulation parameters: $\mathrm{\theta_{incident} \ = \ 0^\circ}$, n\textsubscript{medium} = 1, n\textsubscript{silicon} = 4.2, n\textsubscript{PMMA} = 1.49, dn/dT = - 1.1 $\times$ 10\textsuperscript{-4} K\textsuperscript{-1}, dr/dT = 90 $\times$ 10\textsuperscript{-6} K\textsuperscript{-1}, $T_0$ = 298 K.}
    \label{fig:fig1}
	\end{center}
\end{figure}

Figure~\ref{fig:fig1} shows the photothermal contrast mechanism in dark-field illumination MIP microscopy. A visible laser beam $E_{inc}$ illuminates the sample placed on top of a silicon substrate in the epi-configuration (see Figure~\ref{fig:fig1}a). The incident field scatters off the sample $E_{scat}$ and reflects from the substrate surface $E_{ref}$. The superposition of incident and reflected light constitutes the total driving field of this scattering process. The scattered field is proportional to its optical parameters including refractive index (n) and size (r) along with the illumination wavelength ($\lambda$). The dark-field illumination rejects 10\textsubscript{3} of the reflected light in the collection path such that only the scattered fields reach the detector. A mid-infrared laser beam vibrationally excites the sample owing to the absorption bonds at the IR illumination wavelength. The IR absorption generates heat and increases the sample's temperature by $\Delta T$. This photothermal effect results in a change in the samples refractive index and size which depends on the sample's thermo-optic ($\mathrm{\frac{dn}{dT}}$) and thermal-expansion ($\mathrm{\frac{1}{r}\frac{dr}{dT}}$) coefficients at pre-IR pulse temperature $T$ and hence the scattered field. To obtain the photothermal signal, the scattering difference between IR-on and IR-off states is measured. We refer to IR pulse on and off states as respectively, ``hot'' and ``cold'' frames throughout the manuscript. Figure~\ref{fig:fig1}b shows the normalized radiation profiles of scattering from a 500~nm PMMA bead on the silicon substrate. The zoom-in region indicates a very subtle signal difference ($\sim$ 0.03\% for $\Delta T = 1 K$) between the ``hot'' and ``cold'' states. It is evident that the photothermal radiation profile has a broadened angular distribution compared with the scattering profiles. This suggests a careful treatment of image formation considering imaging optics is required for accurate photothermal signal modeling. With that, theoretical analysis of illumination and collection engineering can be carried out to achieve sensitivity and resolution improvement in the imaging system.

\subsection{\sectionsize Theory}
In dark-field illuminated epi-detection configuration, only the back-scattered light from the particle within the angular range of the objective lens' numerical aperture is collected. The camera captures the resulting scattered field as intensity images ($I_{det} = |E_{scat}|^2$). The IR absorption induced change in refractive index ($\Delta$n) and particle size ($\Delta$r) modifies the scattered field. The scattering field change $\Delta E_{scat} = |E_{scat}^{Hot}|-|E_{scat}^{Cold}|$ is typically three orders of magnitude smaller than the pre-IR pulse scattered amplitude ($\Delta E_{scat} \ll |E_{scat}|$). With this assumption, the photothermal signal $\Delta I_{det}$ can be approximated as follows:

\begin{equation}
    \Delta I_{det} \approx  2|E_{scat}|\Delta E_{scat}
    \label{eq:PT}
\end{equation}

This result shows that $\Delta E_{scat} $ is detected through interference with its own scattered field $E_{scat}$. This is similar to the interferometric detection of photothermal signal which detects $\Delta E_{scat} $ with a reference field $E_{ref}$ such that photothermal signal becomes $2|E_{ref}|\Delta E_{scat}$~\cite{COBRIPhotothermal}. One might expect that both detection schemes could achieve similar sensitivity. Yet, the dark-field photothermal signal has a larger amplitude for a given number of total collected photons due to the fact that the amplitude of $\Delta E_{scat} $ scales with the $E_{scat}$ which is much smaller in the interferometric detection case. The enhancement factor over interferometric detection scales with the ratio between the scattered and reflected fields ($|E_{scat}|/|E_{ref}|$), which determines the signal contrast in such condition~\cite{iSCAT,yurdakulOL,COBRIpupil}. Therefore, one can improve the photothermal signal by reducing the background while keeping the detector at the saturation level. We note that this improvement is only valid for $|E_{scat}| \ll |E_{ref}|$ in the bright-field illumination (see Supplementary 1 Section 1 for details). To obtain a generalized photothermal signal quantification, we use modulation depth which is defined as a fractional change in the scattered intensity ($\Delta I_{det} / I_{det} = 2 \Delta E_{scat} / |E_{scat}| $).  $\Delta E_{scat}$ can be approximated as $\frac {dE_{scat}} {dT} \Delta T$ in narrow temperature intervals. This assumption implies that the photothermal signal scales linearly with the $\Delta T$ for a known specimen. Therefore, one can infer the temperature change distribution of detected particles using \textit{a priori} knowledge of the sample's physicochemical parameters (see Supplementary 1 section 2). 

The scattering-based measurements could bring the system into the shot-noise limited regime where all other noise sources, i.e, electronic and thermal, are negligible if the scattering intensity can saturate the detector at a reasonable exposure time. The shot-noise-limited detection is typically granted for bright-field detection in which a strong reference field enhances the weak scattering signal\cite{iSCAT}. Yet, the particles of interest in this study generate enough scattered photons to bring the camera into the shot-noise-limited regime at a few milliseconds exposure time in the dark-field detection. The noise-floor in a single measurement is then dominated by the photon noise of photoelectrons accumulated at the detector during the integration time. This is a valid assumption for particles that generates enough photons ($N_{det}$) to saturate the detector within a given short exposure time. The shot-noise fluctuation is equal to the standard deviation of the detected photons $\sigma_{photon}= \sqrt{N_{det}}$. In such case, the signal-to-noise ratio (SNR) in the dark-field photothermal signal detection becomes, 
\begin{equation}
    SNR =  \sqrt{2}\Delta E_{scat}
    \label{eq:SNR}
\end{equation}
The $\sqrt{2}$ constant comes from the fact that the noise in hot and cold images are independent of each other and hence the subtracted image noise scales with the $\sqrt{2N_{det}}$. The modulation depth limits the sensitivity in a single shot. From Eq. \ref{eq:SNR}, the modulation depth should satisfy $ \Delta E / |E| >1/ \sqrt{2N_{det}}$ to be detected in a single frame. The maximum $N_{det}$ is bounded by the camera sensor's pixel well depth ($N_{well}$). Therefore, large pixel depth cameras are desirable in shot-noise-limited measurements. Furthermore, the frame averaging is typically employed to increase SNR which is necessary to detect the very subtle scattered field change.

\begin{figure}[!ht]
	\begin{center}
	\includegraphics[width=\textwidth]{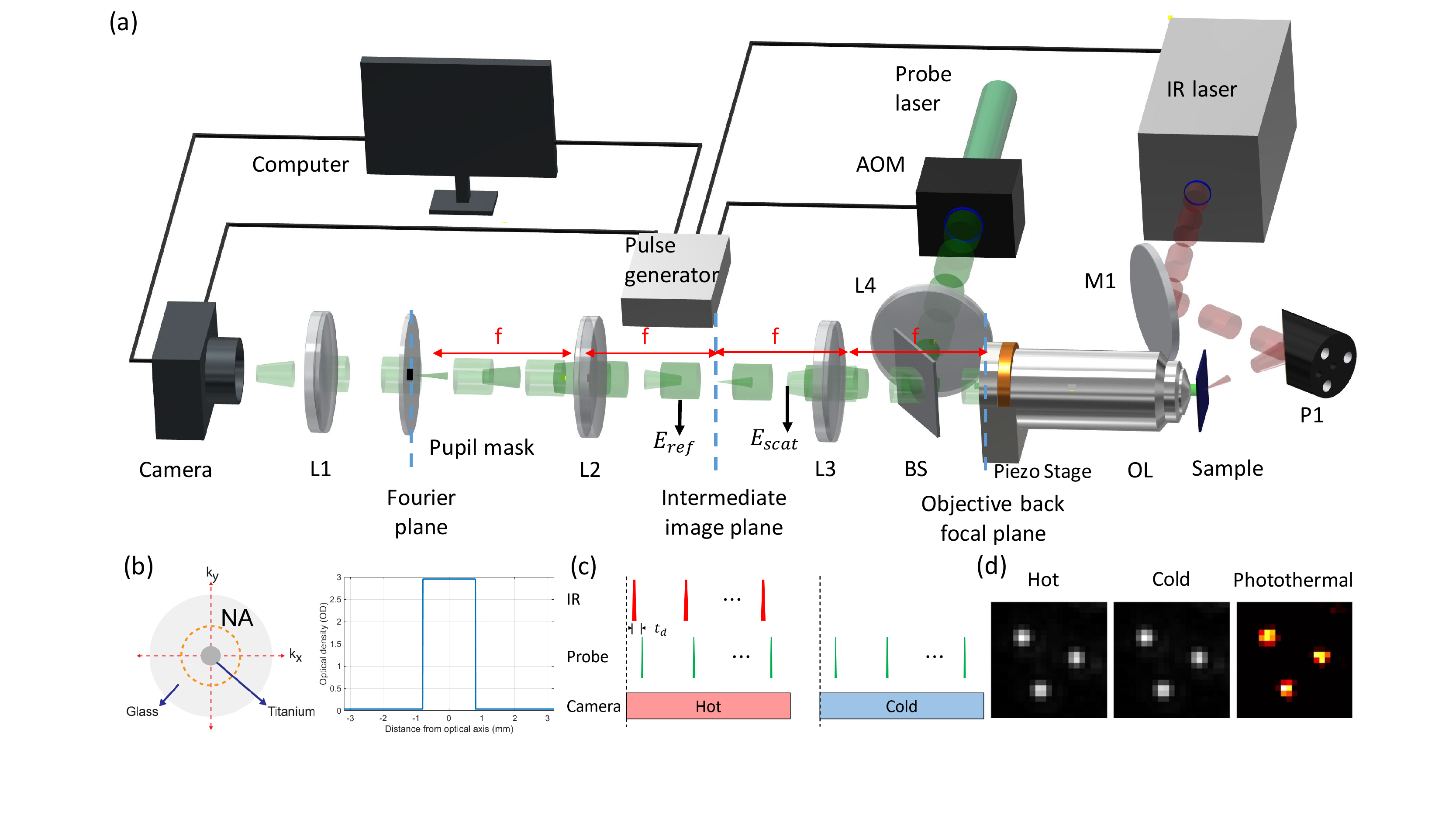}
	\caption{\textbf{Pupil engineering and detection concept in wide-field  MIP system.} (a) Illustration of the experimental setup. AOM: acousto-optic modulator. OL: objective lens. BS: beam-splitter. L1-L4: achromatic doublets. P1: parabolic gold mirror. M1: gold mirror. $\theta$ is the IR incident angle. (b) (Left) The absorptive pupil filter drawing in the objective Fourier plane and (Right) its optical density (OD) cross-section profile within the objective numerical aperture (NA) range. (c) The synchronized acquisition control is triggered by the pulse generator. Timing of pump (the mid-IR) pulses, probe pulses, and the camera frames. $t_d$: time delay between pump and probe pulses. The hot (IR-on) and cold (IR-off) frames. (d) Experimental hot, cold, and photothermal images of 500~nm PMMA beads on the silicon substrate. In the hot image, the IR wavelength is tuned to the 1729~cm\textsuperscript{-1} vibrational peak of the C=O bond. IR power: 6~mW.}
    \label{fig:fig2}
	\end{center}
\end{figure}

\subsection{\sectionsize Pupil engineering for objective-type dark-field illumination}

The experimental setup is illustrated in Figure~\ref{fig:fig2}. The IR beam is provided by a tunable QCL (quantum cascade laser, MIRcat, Daylight solutions). The green pump beam ($\lambda $ = 520~nm central wavelength, $\Delta \lambda$ = 9~nm bandwidth) is obtained by the second-harmonic generation of a quasi-continuous femtosecond laser (1040~nm, $\sim$ 100 fs, 80 MHz, Chameleon, Coherent) using a non-linear crystal. Such a short pulse provides a low-temporal coherence length of $\sim$ 30~µm, yielding nearly speckle-free sample illumination. The femtosecond beam is chopped to pulses (100 ns pulse width) by an acousto-optical modulator (AOM, Gooch and Housego) before entering the non-linear crystal. The optical power after the SHG crystal becomes approximately 30~mW at the continuous AOM operation. The IR beam is weakly focused on the sample from the backside by a parabolic mirror (f = 15~mm, MPD00M9-M01, Thorlabs), illuminating nearly the FOV of 70 µm $\times$ 70 µm. We chose silicon as a substrate for two reasons: (1) silicon is transparent at the IR range and (2) back-scattered light intensity from a particle on the silicon substrate is about 10 times larger compared with a glass substrate (see supplementary Figure~S3). The latter comes from the fact that silicon reflectively is much larger than glass. The strong forward scattering from large dielectric nanoparticles can be collected more on top of a silicon substrate in epi-detection, achieving better collection efficiency given the illumination power. The IR illumination is delivered from the backside of the silicon substrate. The p-polarized IR beam is obliquely incident at a $\theta = 60.3^{\circ}$ angle to increase the transmission rate utilizing Brewster's angle (see Supplementary 1 Section 3 for details). Moreover, the oblique illumination avoids the IR absorption by the objective lens. The probe beam is employed in the K\"{o}hler illumination configuration where the collimated laser is focused on the back focal plane of the objective lens (50$\times$, 0.8 NA, Nikon) by a condenser (f = 75~mm, AC254-075-A, Thorlabs). This provides wide-field plane wave illumination of the sample. Due to the power loss along the optical path and AOM chopping (1:50), $\sim$ 0.6~mW probe power illuminates the sample across the FOV of 190 µm $\times$ 190 µm. The objective lens is mounted on a piezo stage (MIPOS 100 SG RMS, Piezosystem Jena) for fine focus adjustment. Besides, the piezo scanner eliminates the need for defocus adjustment of the IR beam, since the sample z position remains unchanged with respect to the IR focus plane. 

The objective-through dark-field illumination is implemented through pupil engineering at the objective pupil's conjugate plane. To achieve this, the objective pupil is relayed by a unit magnification 4f system that uses two identical achromatic doublets (f = 100~mm, AC508-100-A, Thorlabs). As shown in Figure~\ref{fig:fig2}a, the left focal plane of this 4f system (Fourier plane) becomes conjugate to the objective pupil. Since the back-reflected light from the substrate is refocused at the objective back pupil, the reflected light at the conjugate plane becomes accessible. This approach is based on the Unlu's group earlier study in which contrast-enhanced interferometric detection is demonstrated by attenuating a small fraction ($\sim$ 20:1) of the reflected light~\cite{oguzhanOptica}. To enable the dark-field detection, we instead block a large amount (1000:1) of the reflected light by a custom fabricated pupil mask placed into this Fourier plane. The pupil mask is depicted in Figure~\ref{fig:fig2}b. There is a dot blocker with a diameter of 1.6~mm at the center of the mask. This only filters the low-NA portion (0.2 NA) centered around the objective optical axis. The center block dot;s diameter is empirically determined by considering the amount of collected scattered photons as well as the alignment difficulty. The blocker blocks 6\% of the pupil while passing a large fraction of the collected scattered light. The supplementary Figure~S5 shows the block diameter as a function of the collected power for both intensity-only and photothermal cases. According to simulation results, we estimate 83.7\% collection efficiency for the 500~nm beads on a silicon substrate. Therefore, the pupil mask can provide dark-field illumination of wavelength size particles while maintaining the detector at shot-noise-limit operation. As we pointed out earlier, the photothermal effect broadens the angular distribution of the radiation, yielding a lower directivity compared to the DC signal. Consequently, the collected photothermal power is 89\% for the 1.6~mm blocker, which has 5.3\% higher efficiency than the DC case. One can also choose a smaller pupil diameter as long as the blocker overfills the back-reflected light at the Fourier plane. In ideal conditions, the minimum blocker size is equal to the relay magnification times the focused/imaged beam size at the objective back pupil. In practice, the beam divergence, optical aberrations, and system misalignment incur a larger beam size. Furthermore, we expect better collection efficiency for smaller particles since they have much broadened angular distribution owing to the Rayleigh scattering~\cite{mieScattering}. This result also indicates that photothermal modulation depth at high scattering angles becomes larger. Besides, our method can be implemented on most of the standard bright-field objectives which offer a wide range of availability for different applications. This obviates the need for high-cost special objectives which have the dark-field ring at their back-pupil. The dark-field ring blocks the high-NA part of the objective, reducing the attainable resolution. Therefore, our technique does not possess challenges associated with the dark-field objectives for epi-detection. In the implementation, an absorptive material, Titanium, is deposited at the center of optic quality quartz. The titanium thickness is about 80~nm, providing an optical density of $\sim$3. This is almost opaque ($\sim$ 0.1\% transmission) compared to the $\sim$ 80\% for the glass region. We note that there is still a small fraction of the reflected light reaching the detector, allowing for quasi-darkfield illumination. After the pupil mask, the dark-field image is formed on a CMOS camera (BFS-U3-17S7, FLIR, dynamic range of 72.46 dB, dark noise of 22.99 e$^-$) by a tube lens (f = 200~mm, TTL200-A, Thorlabs). The camera region-of-interest is cropped to match the IR spot size at the sample. The camera exposure time during the experiments is set to a level that nearly saturates the camera pixels. In summary, our dark-field illumination scheme provides robust, simple, and low-cost background suppression for contrast enhancement in the epi-detection arrangement.

\begin{figure}[!t]
	\begin{center}
	\includegraphics[width=\textwidth]{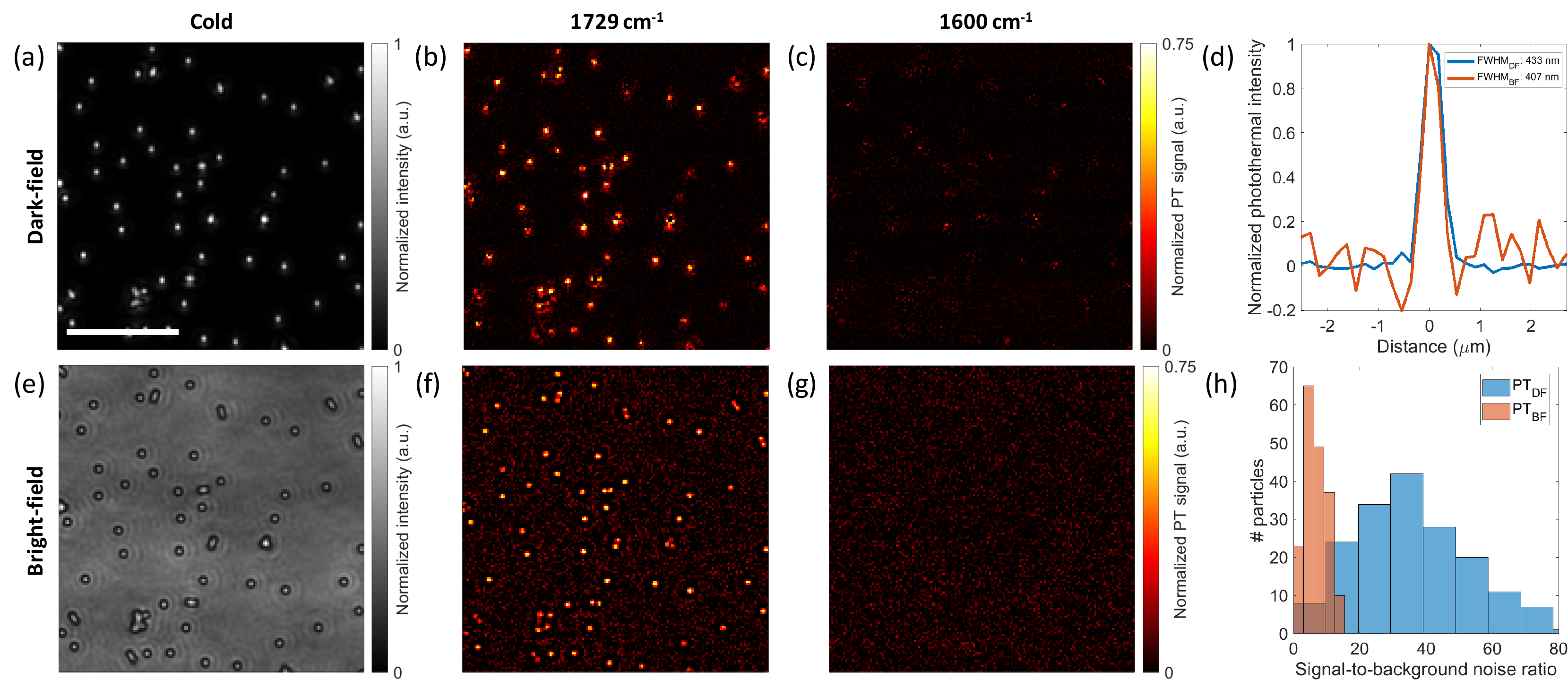}
	\caption{\textbf{Comparison between dark-field MIP and bright-field MIP imaging of 500~nm PMMA beads on silicon substrate.} (a) Dark-field cold image. (b) Dark-field photothermal image at C=O absorption peak (1729~cm\textsuperscript{-1}). (c) Dark-field photothermal image at off-resonance 1600~cm\textsuperscript{-1}. (e-g) Corresponding bright-field images of the same FOV. (d) Cross-section of a selected bead in (b) and (f). (h) Histograms of the signal-to-background-noise ratios calculated from (b). The images are cropped for better visualization by a factor of 2.7, showing fewer beads in the images. Photothermal image acquisition time: 5 s. IR power: 6~mW @ 1729~cm\textsuperscript{-1}, 11.7~mW @ 1600~cm\textsuperscript{-1}. Scale bar: 10 µm.}
    \label{fig:fig3}
	\end{center} 
\end{figure}

\subsection{\sectionsize Time-gated virtual lock-in camera detection}
Photothermal images are acquired by using the virtual lock-in camera detection~\cite{Yeran2019}. Figure~\ref{fig:fig2}c shows the synchronization control of the system. A pulse generator (Emerald Pulse Generator, 9254-TZ50-US, Quantum composers) generates the master clock signal at 200 kHz and externally triggers the QCL, AOM, and CMOS camera to synchronize the IR pulses, probe pulses, and camera exposure. The pulse generator has a division function that enables separately controlling the duty cycle, frequency, amplitude, delay, and width at each output channel. The output waveforms are set to pulse wave. The time delay ($t_d$) between the IR and visible pulses is controlled to measure transient photothermal response via a time-gated pump-probe approach~\cite{phase2019}. We used the AOM channel at the normal mode without changing the duty cycle. The QCL and camera channels are set to the duty cycle mode to create hot and cold frames. The QCL pulse train is chopped electronically at 50\% duty cycle, that is, every other 1000 IR pulses on and off. Similarly, the camera channel duty cycle is set to 2\% to achieve a 400 Hz frame rate at the region of interest. The camera readouts ``hot'' and ``cold'' frames sequentially and streams to the computer via USB. The pre-determined frame averaging is processed in real-time for computational efficiency, significantly reducing the memory requirements at a large number of frames ($O (N) \rightarrow O (1)$), where N is the frame number). To avoid probe laser intensity fluctuations, each image is normalized by the average intensity at a predetermined reference region which is obtained by directly reflecting the split illumination beam to the lower left of the FOV. During the image acquisition, we set the odd- and even-numbered frames as ``hot'' and ``cold'' states, respectively. This allows us to extract photothermal signal sign which has size dependency detailed in the results sections. The photothermal image is then obtained by subtracting the ``hot'' and ``cold'' images as shown in Figure~\ref{fig:fig2}d. A custom-written Python code is developed to automatically control pulse delay, piezo scanner, image acquisition, and processing (detailed in Supplementary 1 section 5).

\subsection{\sectionsize Experimental verification of contrast enhancement}

Proof-of-principle experiments for contrast enhancement are demonstrated on 500~nm Polymethyl methacrylate (PMMA) beads. The PMMA beads present an ideal model for the system characterization since they resemble the particle size and dielectric ($n \approx 1.49$) characteristics of bacteria used in our study. Figure~\ref{fig:fig3} compares the dark-field and bright-field imaging results. The dark-field illumination is achieved by placing the blocker mask at the pupil conjugate plane detailed in the instrumentation section. For a fair comparison, bright-field imaging results are obtained by the same setup at the same conditions without the pupil mask. The exposure time in both cases is adjusted to bring the camera to the saturation level. As shown in Figure~\ref{fig:fig3}a,e, the dark-field illumination provides background-free DC imaging while the bright-field image has a non-zero background caused by the reflection from the silicon substrate. Although the non-zero background in the DC images can be canceled by subtraction, it still contributes to shot-noise which limits the maximum attainable SNR from a single frame. Figure~\ref{fig:fig3}f shows the background shot-noise clearly, which degrades the visibility of the PMMA beads. In the dark-field photothermal image (Figure~\ref{fig:fig3}b), the background shot noise is significantly eliminated. The cross-section profiles of the bead shown in Figure~\ref{fig:fig3}d further emphasize this significant background noise suppression. The resolution capability of the imaging system can be also calculated from these cross-sections. The full-width-half-maximum (FWHM) of a 500~nm bead is 433~nm for the dark-field imaging and 407~nm for the bright-field imaging. This result shows that our dark-field illumination approach can achieve the nearly same resolution as in the bright-field illumination. One can expect that the dark-field FWHM would be smaller due to the fact that the interferometric signal in the bright field case has a broader point spread function (PSF). However, the 500~nm PMMA bead scattered intensity has still a contrast level comparable to the interferometric signal, providing a sharper PSF than that of the interference term. We anticipate that the dark-field detection will provide better resolution for much smaller nanoparticles (100~nm PMMA beads) which have much weaker contrast in the bright-field imaging. After a Gaussian deconvolution with the particle size, we obtained 353~nm lateral resolution which is close to the theoretical $ \frac{\lambda}{\textrm{2NA}}$ = 325~nm  resolution value. The experimental FWHM is slightly larger than the theoretical resolution value because the 500~nm bead is not small enough to be approximately treated to be a point source. To obtain a more quantitative analysis metric, the signal-to-background-ratio histograms of 190 PMMA beads are compared in Figure~\ref{fig:fig3}h. The median signal-to-background-noise ratio for the dark-field case is about 6 times larger than the bright-field case, reaching up to 100. The signal-to-background-noise improvement of this dark-field illuminated MIP imaging system has been demonstrated for high-throughput chemical imaging of wavelength size single particles. 

\begin{figure}[!t]
	\begin{center}
	\includegraphics[width=\textwidth]{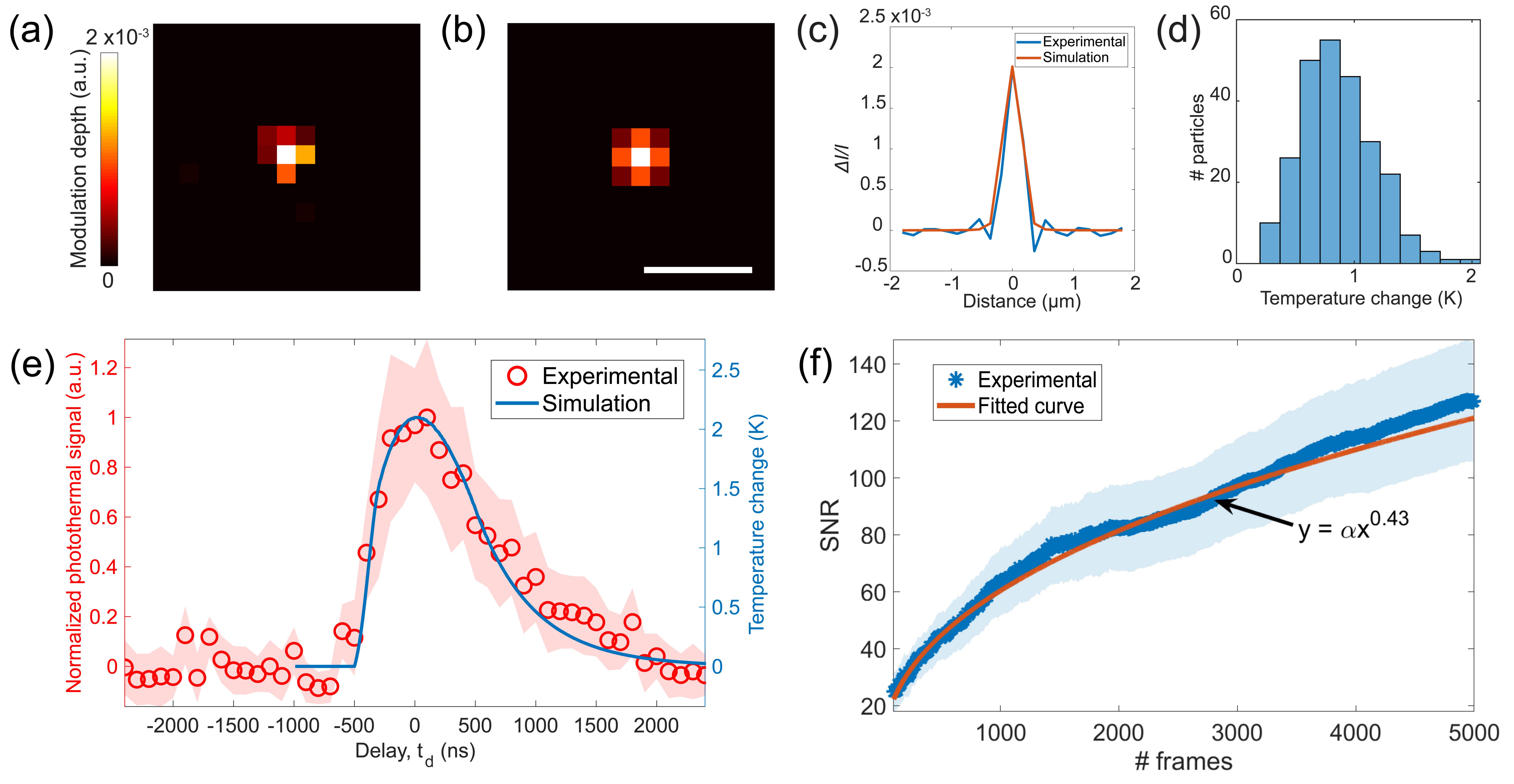}
	\caption{\textbf{Experimental validation of photothermal image formation modeling.} (a) Experimental and (b) simulated photothermal image of a 500~nm PMMA bead at 1729~cm\textsuperscript{-1}. (c) Modulation depth ($ \Delta I / I $) cross-section profiles in \textbf{a-b}. (d) Temperature change ($\Delta T $) histogram of the detected 500~nm PMMA beads in Figure~\ref{fig:fig3}. Temperature change is calculated at each bead's peak contrast using the linear relationship with the modulation depth. (e) Experimental and simulated transient temperature response for 56 particles. Temperature decay time constant is 915.6 ns. (f) Photothermal image SNR calculated at different number of frames averaging. Experimental data is fit to an exponential function $y = \alpha x^n$ with n = 0.43. IR power: 6~mW @ 1729~cm\textsuperscript{-1}. Scale bar: 1 µm.}
    \label{fig:fig4}
	\end{center}
\end{figure}

\subsection{\sectionsize Experimental validation of theoretical calculations}
The image formulation framework detailed in the Methods section is verified in two steps using the experimental photothermal image of a 500~nm PMMA bead. We first calculate the modulation depth from the BEM simulation using the PMMA's optical and thermal coefficients at $\Delta T \ = \ 1 \ K$. Since the modulation depth can be linearly related to the small temperature changes, $\Delta T$ of the PMMA beads can be retrieved from the experimental results (see Supplementary 1 Section 2 for details). The experimental photothermal modulation depth image of a 500~nm PMMA bead on the silicon substrate is shown in Figure~\ref{fig:fig4}a. The modulation depth is calculated as the ratio between photothermal image and peak contrast value at the cold state. Figure~\ref{fig:fig4}b is a simulation photothermal modulation depth image which is scaled to the same maximum value in Figure~\ref{fig:fig4}a. The cross-section profiles in Figure~\ref{fig:fig4}c show consistency between the experimental and simulated results. We then calculate the $\Delta T$ histogram for all PMMA beads Figure~\ref{fig:fig3}b as shown in Figure~\ref{fig:fig4}d. The maximum temperature rising across the FOV is calculated as $\sim$ 2 K which is consistent with the COMSOL simulations (see Supplementary Section 4 for details). 

The simulated temperature rising and the experimental photothermal signal versus delay scan of 56 individual PMMA beads with a 500~nm diameter is shown in Figure~\ref{fig:fig4}e. For each specific delay value, the photothermal signal is proportional to the integrated temperature change within the time window of the probe pulse, which has a 200 ns pulse width as introduced in the previous section. In other words, the curve shape of the experimental delay scan is a convolution of the simulated temperature curve with the 200 ns probe pulse. The experimental delay scan curve is not distorted too much compared to the simulation, which means the 200 ns pulse width is short enough to probe the highest temperature change. We point out that the transient response curves depend on sample size and IR pulse shape. Considering the pulse shape and particle size in this study, the time delay is carefully determined to obtain the maximum photothermal signal during the experiments. The photothermal images are then acquired using the optimized delay scan value that corresponds to the highest photothermal signal.

The photothermal signal scales linearly with the probe power in the shot-noise limit. A maximum SNR in a signal frame is then can be achieved at the camera saturation level where the shot-noise-limited detection is granted. The minute contrast change as a result of the photothermal effect can be detected through multiple frames averaging which reduces the noise floor by a factor of the square root of the number of averaged frames. Supplementary Figure~S12 demonstrates the shot-noise limited operation for our camera and noise floor reduction by frame averaging. Figure~\ref{fig:fig4}f further shows the noise analysis of 56 PMMA beads over a different number of frames averaging $N$. The noise is calculated as the standard deviation of the photothermal image background nearby the particles. The exponential fit to the experimental SNR values is found to be $SNR \propto N^{0.43}$. The slight variation from the theoretical value of 0.5 could be attributed to the mechanical noise in the imaging system. The photothermal image results at different frame averaging have been also shown in supplementary Figure~S13. Furthermore, we obtain a photothermal spectrum of 20 individual beads in the supplementary Figure~S14. The IR wavenumber is scanned from 1750 to 1400~cm\textsuperscript{-1} with a step size 1~cm\textsuperscript{-1}, requiring nearly 29 minutes of total acquisition time. The image acquisition and hyperspectral analysis are detailed in the supplementary information. The spectra show distinctive absorption peaks around the carboxyl group and C-H bonds with high spectral fidelity. Our experimental results are consistent with those obtained from the point-scan MIP imaging reported in~\cite{IRaman}. Overall, our characterization has shown the potential of dark-field MIP microscopy to facilitate highly sensitive hyperspectral imaging of wavelength size particles.

\begin{figure}[!t]
	\begin{center}
	\includegraphics[width=.6\textwidth]{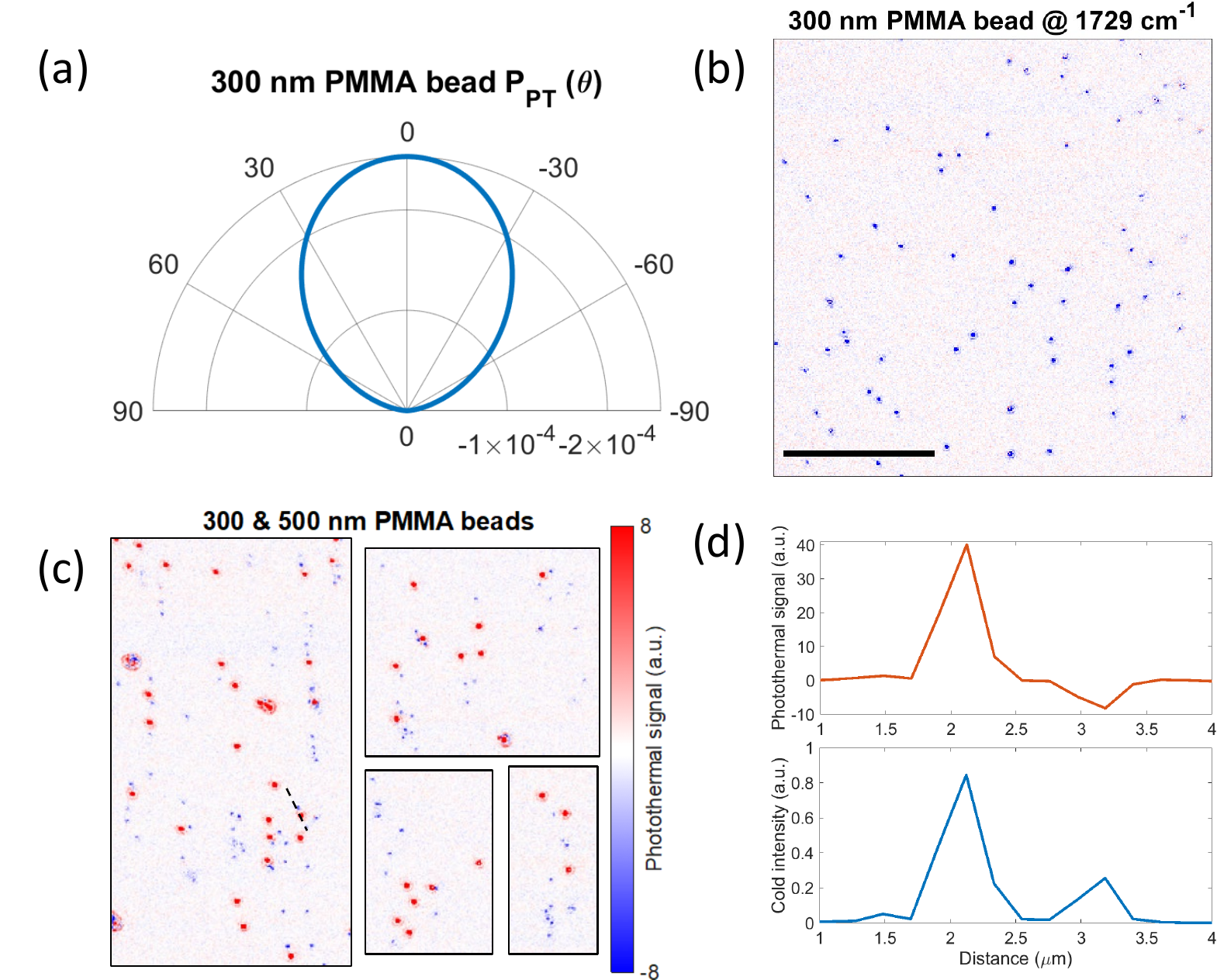}
	\caption{\textbf{Size dependence of photothermal signal sign.} (a) Photothermal scattered intensity polar plot of 300~nm PMMA bead on silicon substrate. Signal is normalized by the maximum intensity value in the cold state. Simulation parameters are the same as in Figure~\ref{fig:fig1}. (b) Photothermal image of 300~nm PMMA beads. (c) Cropped photothermal images of 300 and 500~nm PMMA beads mixture. (d) Cross section profile of the blue dash line in (c) (top) and the corresponding profile in the cold image (bottom). The IR wavelength is tuned to 1729~cm\textsuperscript{-1}. Photothermal image acquisition time: 25 s (5000 frames). IR power: 6~mW @ 1729~cm\textsuperscript{-1}. Scale bar: 20 µm. }
    \label{fig:fig5}
	\end{center}
\end{figure}

\subsection{\sectionsize Size dependence of photothermal contrast}
We next investigate the size dependence of photothermal signal using 300~nm PMMA beads in diameter. Figure~\ref{fig:fig5}a shows the scattering intensity polar plot of the photothermal signal from a PMMA bead of 300~nm diameter. The radiation spreads more uniformly across the angles, indicating a lower directivity compared with the 500~nm PMMA radiation profile in Figure~\ref{fig:fig1}b. This stems from the well-known Mie-scattering fact that far-field scattering angular distribution has a strong dependency on particle size. The directivity of the radiation is inversely proportional to the scatterer's size. We note that both polar plots are normalized by the maximum intensity value at the cold state. More importantly, unlike the positive contrast in 500~nm PMMA beads, 300~nm PMMA beads have negative photothermal contrast. This could be explained by the self-interference between the back-scattered and forward scattered fields from the same particle. This self-interference of scattered fields occurs since the forward scattered fields from particles reflect from the substrate surface. The forward scattered light becomes less dominant for the smaller particles due to the Mie-scattering phenomena. Therefore, the photothermal contrast sign flip is likely to happen when the amplitude of the forward scattered field decreases. Another reason for the photothermal signal sign change comes from the fact that the refractive index and thermal expansion of PMMA beads counteract each other due to the opposite sign\cite{Zhongming2017}. Interestingly, the photothermal signal contribution from these effects could cancel each other at a certain size depending on the magnitude of these thermal coefficients as well as the aforementioned phase relation between forward and backward scattered fields. Therefore, certain particles of interest depending on their size and material properties could be invisible which could eventually limit the certain applications in scattered-based MIP techniques~\cite{Zhongming2017}. Due to the complexity of such analysis with a closed-form solution, we numerically investigate the size-dependent photothermal signal from a size range of PMMA beads on the silicon substrate in Supplementary Figure~S15. The inversion for the PMMA bead occurs around 350~nm in diameter. We experimentally verified the sign inversion using a 300~nm PMMA bead sample. Figure~\ref{fig:fig5}b shows the photothermal image of the 300~nm PMMA beads with a signal-to-background-noise ratio of 44 at the frequency of 1729~cm\textsuperscript{-1}. The cold state and off-resonance images at 1600~cm\textsuperscript{-1} are provided in supplementary Figure~S13. We further cross-validate our theory using a mixed 300 and 500~nm PMMA beads sample (Figure~\ref{fig:fig5}c). Depending on the particle size, the photothermal signals from different particles yielded positive or negative contrast in a single FOV. Our experimental findings show great agreement with the theoretical predictions. The cross-section profiles in Figure~\ref{fig:fig5}d demonstrate a clearer comparison of scattered intensity and contrast flip. Due to the sixth power dependence (r\textsuperscript{6}) of the scattered signal, 300~nm beads yield lower DC contrast. Moreover, the temperature rise for smaller particles becomes less due to a faster thermal decay in the orders of a few hundreds of nanoseconds, requiring shorter pump and probe pulse widths (Supplementary Figure~S16). As a result, lower photothermal SNR for 300~nm beads is obtained. To obtain high SNR images with good data fidelity in these proof-of-concept experiments, the number of frame averaging is increased by five-fold to 5000. 

\begin{figure}[!t]
	\begin{center}
	\includegraphics[width=.9\textwidth]{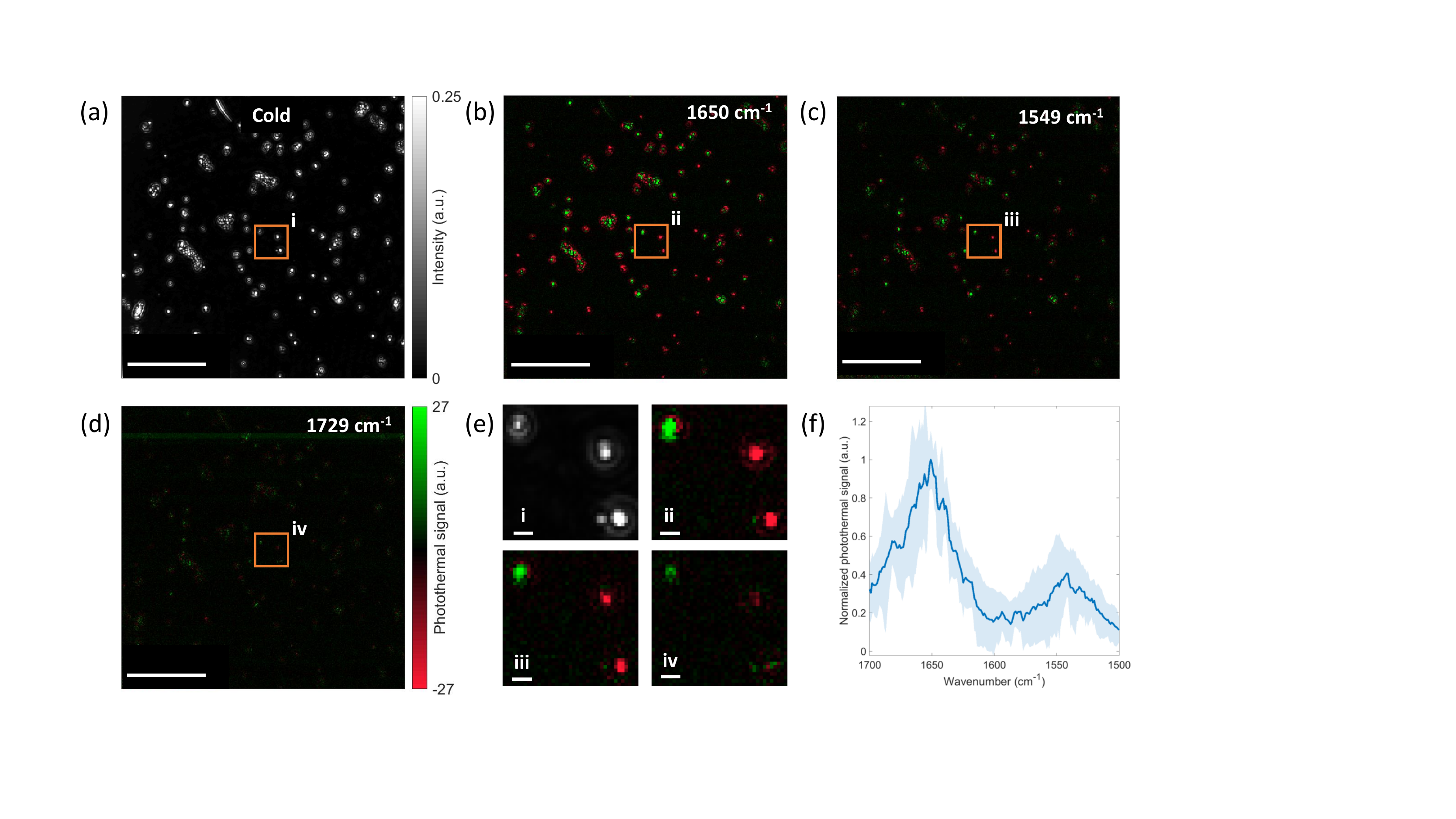}
	\caption{\textbf{Multispectral dark-field MIP imaging of \textit{S. aureus}.} (a) Dark-field cold image of \textit{S. aureus}. (b-d) Dark-field photothermal images of \textit{S. aureus} at specific wavenumbers for different chemical bonds. (e) Dark-field MIP spectrum of \textit{S. aureus}. Background standard deviation in photothermal images is around 0.84. Photothermal image acquisition time: 5 s (1000 frames). Total spectral scan time: $\sim$21 mins. IR power: 6~mW @ 1729~cm\textsuperscript{-1}, 11.9~mW @ 1650~cm\textsuperscript{-1}, 8.7~mW @ 1549~cm\textsuperscript{-1}. MIP spectrum is normalized by the IR power. Scale bars in (a-d) is 20 µm, and in zoom-in areas (e) are 1 µm.}
    \label{fig:fig6}
	\end{center}
\end{figure}

\subsection{\sectionsize Fingerprinting single bacteria}

To demonstrate the utility of our method on biological specimens, we study two bacteria species with various sizes and shape distribution. The bacteria are directly immobilized on the silicon substrate at room temperature (Supplementary 1 Section 5). We first imaged spherical \textit{S. aureus} bacteria in the fingerprint region. Figure~\ref{fig:fig6}a demonstrates cold image in which the single \textit{S. aureus} cells appear to be round. The intensity variation across the bacteria indicates size differences of \textit{S. aureus} cells. When IR frequency is tuned to the amide $\mathrm{\uppercase\expandafter{\romannumeral1}}$ band at 1650~cm\textsuperscript{-1}, which is a characteristic band in proteins, the bacteria show high-contrast photothermal signal with a signal-to-background-noise ratio of 93 (Figure~\ref{fig:fig6}b). The photothermal image indicates the rich protein in \textit{S. aureus} cells. The photothermal contrasts from different bacteria cells show not only amplitude variation due to scattered intensity differences but also negative or positive contrast depending on their size. These observations are consistent with the one obtained from the PMMA beads above. We then pinpointed another major protein band of amide $\mathrm{\uppercase\expandafter{\romannumeral2}}$ at 1549~cm\textsuperscript{-1} (Figure~\ref{fig:fig6}c). In contrast to amide $\mathrm{\uppercase\expandafter{\romannumeral1}}$, amide $\mathrm{\uppercase\expandafter{\romannumeral2}}$ generates lower photothermal contrast owing to the weaker absorption. When IR is tuned 1729~cm\textsuperscript{-1}, associated with C=O bond which is abundant in lipids, a very weak contrast is observed as a result of low lipid content in the \textit{S. aureus} (Figure~\ref{fig:fig6}d). Furthermore, Figure~\ref{fig:fig6}e shows spectrum of 12 individual bacteria from 1700 to 1500~cm\textsuperscript{-1} with a step size of 1~cm\textsuperscript{-1}. Our results show agreement with the previously reported study using the point-scan MIP~\cite{IRaman}. With these results, single bacteria fingerprinting in a wide-field imaging system has been demonstrated. Owing to the wide-field detection, simultaneous photothermal detection of tens of bacteria is achieved.

\begin{figure}[!t]
	\begin{center}
	\includegraphics[width=.9\textwidth]{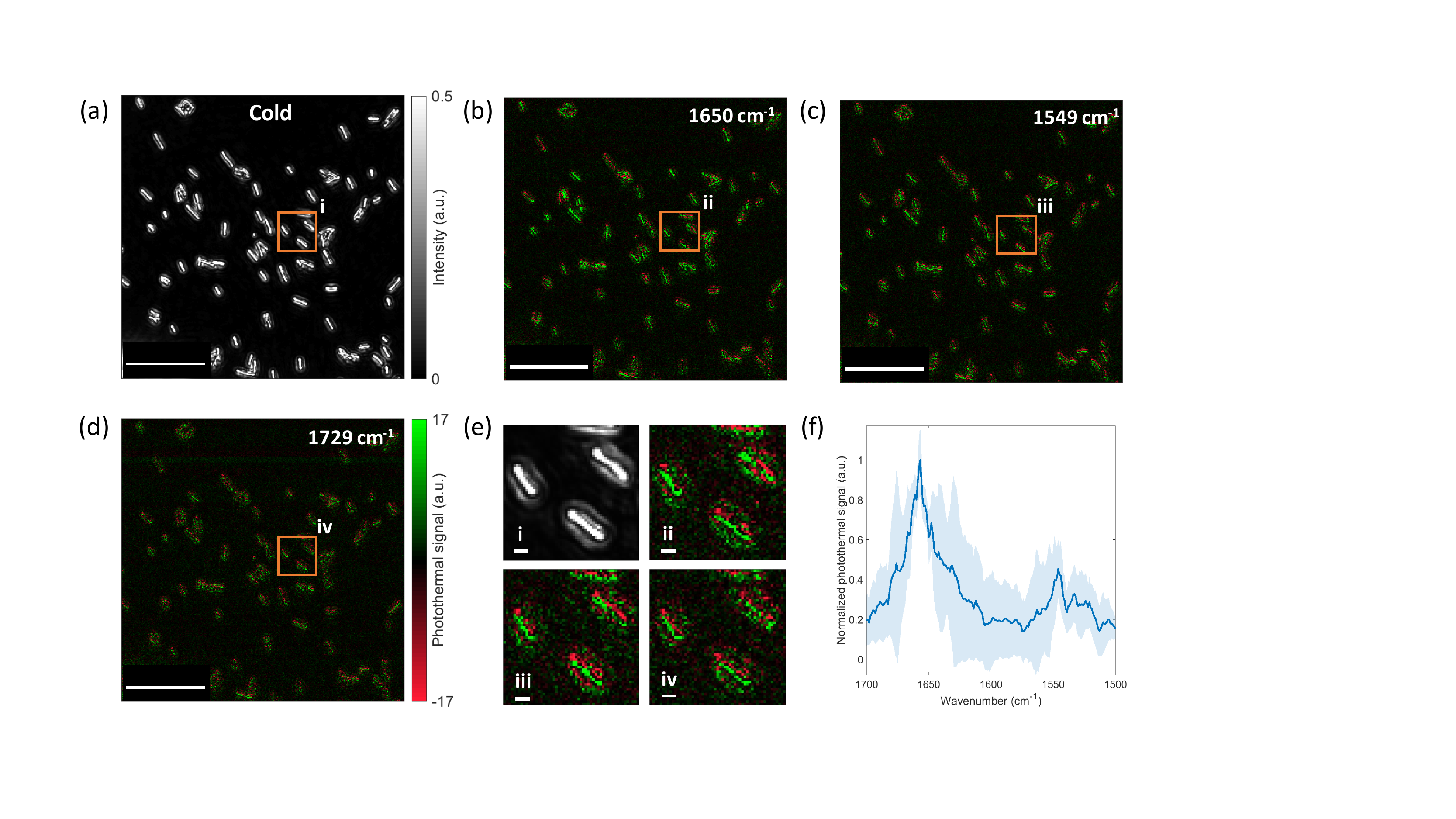}
	\caption{\textbf{Multispectral dark-field MIP imaging of \textit{E. coli}.} (a) Dark-field cold image of \textit{E. coli}. (b-d) Dark-field photothermal images of \textit{E. coli} at specific wavenumbers for different chemical bonds. (e) Dark-field MIP spectrum of \textit{E. coli}. Background standard deviation in photothermal images is around 0.99. Photothermal image acquisition time: 5 s (1000 frames). Total spectral scan time: $\sim$21 mins. IR power: 6~mW @ 1729~cm\textsuperscript{-1}, 11.9~mW @ 1650~cm\textsuperscript{-1}, 8.7~mW @ 1549~cm\textsuperscript{-1}. MIP spectrum is normalized by the IR power. Scale bars in (a-d) is 20 µm, and in zoom-in areas (e) are 1 µm.}
    \label{fig:fig7}
	\end{center}
\end{figure}

Next, we demonstrated hyperspectral characterization of rod-shaped \textit{E. coli} bacteria in the fingerprint region. Figure~\ref{fig:fig7}a shows scattered intensity image of \textit{E. coli}. Unlike the \textit{S. aureus} bacteria, the intensity variation across the many \textit{E. coli} bacteria is small. This suggests that the diameter of each \textit{E. coli} bacterium is almost monodisperse. Figure~\ref{fig:fig7}7b-d compare photothermal images of amide $\mathrm{\uppercase\expandafter{\romannumeral1}}$, amide $\mathrm{\uppercase\expandafter{\romannumeral2}}$, and off-resonance bands at frequencies of 1650~cm\textsuperscript{-1}, 1549~cm\textsuperscript{-1}, and 1729~cm\textsuperscript{-1} respectively. We obtained about four-fold lower photothermal signal compared with the large \textit{S. aureus} cells due to the smaller diameter of \textit{E. coli} bacteria. The photothermal contrast at the bacteria center is always positive owing to their uniform diameter distribution. We further obtain spectra of 10 bacteria (Figure~\ref{fig:fig7}e) from 1700 to 1500~cm\textsuperscript{-1} with a step size of 1~cm\textsuperscript{-1}. The \textit{E. coli} and \textit{S. aureus} spectra show different curves across the amide $\mathrm{\uppercase\expandafter{\romannumeral1}}$ and  $\mathrm{\uppercase\expandafter{\romannumeral2}}$ bands. These observations lead to potential applications in high-throughput single bacteria characterization and classification. 

\section{\sectionsize CONCLUSION}
We have introduced a contrast-enhancement method in wide-field MIP microscopy with pupil engineering that enables dark-field illumination through a bright-field high-NA objective. We achieved three orders of magnitude background suppression in DC images while maintaining diffraction-limited lateral resolution. Our dark-field illuminated MIP technique improved the signal-to-background ratio at least 6-fold over a large FOV of 70 by 70~µm that allows for high-throughput and sensitive chemical imaging of hundreds of sub-micron particles simultaneously. We showed the transient temperature response of 500~nm PMMA nanoparticles at sub-200~ns temporal resolution via a time-gated pump-probe approach implemented on a commercially available CMOS camera. We experimentally demonstrated hyperspectral imaging of single \textit{S. aureus} and \textit{E. coli} bacterium in the fingerprint region with high SNR and high spectral fidelity in both experiments. Single sub-micron bacterium fingerprinting in a wide-field manner has been demonstrated. Moreover, a comprehensive analytical model for photothermal contrast mechanism and image formation was described and experimentally validated. This physical model opens up further advancements in photothermal microscopy through careful signal characterization with the optical system's constituent specifications, including point spread function, illumination and collection functions, and substrate.

We note that an independent study using a similar background suppression concept has been recently reported during the peer-review process of our work~\cite{ADRIFT}. Toda \textit{et al.} used the dark-field scheme to expand the dynamic range in mid-IR photothermal quantitative phase imaging. Dark-field detection in transmission mode is realized by attenuating the strong bright-field illumination at the pupil conjugate plane by a dot blocker. Dynamic range expansion of quantitative phase imaging of micrometer-scale specimens such as 5 µm silica beads and COS cells has been successfully demonstrated. This phase imaging study used specimens that are much larger than the illumination wavelength, requiring a low-refractive-index difference between the surrounding medium. Therefore, single nanoparticle detection sensitivity is still challenging. Despite the methodical similarity, detection mechanism, imaging configuration, and sensitivity draw major differences. First, detection principles of sub-diffraction-limited nanoparticles rely on elastically scattered light. Second, we demonstrate 10-fold enhanced scattered light detection using silicon substrate in epi-illumination configuration which is critical for successfully suppressing the background without compromising from illumination power. Third, our scope of work is to study nanoscale specimens with high optical detection sensitivity down to a single 300~nm nanoparticle.


The dark-field illumination could be a natural choice for specimens that scatters enough photons to saturate the camera. There are two drawbacks to this approach. First, it does not benefit large particles that can already generate high image contrast in bright-field illumination or dense specimens such as cells and tissues that generates large background signal. Therefore, our method is well suited for individual particles sparsely distributed across the substrate. This will be of significant importance for enabling chemical characterization at the single-particle level for understanding variations in the diverse nanoparticle populations. The other limitation is the strong (sixth-power) size dependence of the scattering intensity (r\textsuperscript{6}) which significantly drops for sub-100~nm nanoparticles. Therefore, the applications are limited to the wavelength scale of individual particles to maintain shot-noise limited detection. To push the sensitivity limit down to a single virus or exosome, common-path interferometric detection of the scattered light in tandem with pupil engineering could be a path forward in future wide-field MIP studies~\cite{oguzhanOptica,yurdakulOL}. This will not only improve the imaging system's sensitivity but also the lateral resolution down to 150~nm~\cite{yurdakulAN}. Together, our theoretically supported wide-field MIP microscopy technique could bring exciting applications in life sciences.

\section{\sectionsize MATERIALS AND METHODS}

\subsection{\sectionsize Image field calculations}
To accurately characterize the photothermal contrast mechanism, we developed an analytical model considering imaging optics and system parameters. We employ image field representation of optical fields that provides better means for physical optical system simulations. Our model is built upon the previously developed theoretical framework for interferometric scattering calculations from an arbitrary shape and size particle near a substrate~\cite{SevenlerBOE} and extends to the photothermal signal. The photothermal imaging simulation is split into two steps: (1) numerical evaluation of far-field scattered field from a particle and (2) calculating image fields using diffraction integrals. To do so, we first define the system geometry including the substrate, medium, and particle dielectric functions as well as the illumination wavelength ($\lambda$). The vectorial scattered fields at the infinity ($\mathbf{E_{scat,\infty}}$) are then calculated using the metallic nanoparticle boundary element method (MNPBEM) toolbox~\cite{MNPBEM}. This toolbox numerically solves full Maxwell's equations for the dielectric environment in which the particle and surrounding medium have homogeneous and isotropic dielectric functions. In calculations, it utilizes the boundary element methods (BEM)~\cite{BEM} which is a computationally efficient approach for simple geometries. It should be noted that MNPBEM accounts for the substrate effect on internal and driving electric fields using Green's functions. This is very important for accurate analysis of the total back-scattered field considering the reflections from the surfaces. After numerically calculating the far-field scattered field, we perform image formation integrals using angular spectrum representation (ASR) of vectorial electric fields. The ASR framework has been a powerful tool for a rigorous and accurate description of the field propagation in the homogeneous media~\cite{novotnyBook}. The electric field distribution at the image plane can be explained by the superposition of the far-field scattered fields as follows: 

\begin{equation}
    \mathbf{E_{scat}}(x,y,z) = A_0\frac{j}{2\pi} \iint\limits_{\sqrt{k_x^2+k_y^2} \leq k_{NA}} \frac{1}{k_z}\mathbf{E_{scat,\infty}}\left(\frac{k_x}{k},\frac{k_y}{k}\right) e^{j(k_xx+k_yy\pm kz_z)}\,dk_x\,dk_y
    \label{eq:ASRIntegral}
\end{equation}

\noindent where $A_0$ is scaling factor associated with the far-field calculations at the infinity, $k = \lambda/2\pi$ is the wavevector, and $k_z = \sqrt{k^2-k_x^2-k_y^2}$ is the wavevector along the optical axis $z$. The integral limits impose filtering pupil function defined by the objective NA. Therefore, the scattered radiation profile has of great importance for signal calculations. The scattered light intensity is then calculated at the camera plane by taking magnitude square of the image field. To incorporate the photothermal effect into the model, the same steps are iterated after updating the particle size and refractive index using the thermo-optic~\cite{PMMAthermooptic} and thermal-expansion coefficients~\cite{thermalExpansion} explained above. For example in Figure~\ref{fig:fig1}b, the simulation geometry is defined for a 500 PMMA bead (n = 1 .49)~\cite{PMMArefractiveindex} placed on top of a silicon substrate (n = 4.2)~\cite{siliconRefractive}. We set the imaginary part of the silicon refractive index to zero since it is negligibly small compared with the real part at the illumination wavelength ($\lambda$ = 520~nm). We assumed plane wave illumination from above. This is a valid approximation for the nearly collimated sample illumination in the experiments. To speed up the successive simulations, reflected Green's functions are pre-calculated and stored in the memory. A similar approach has been taken to calculate the photothermal signals for different particle sizes.
    
\subsection{\sectionsize Photothermal effect simulations}
The analytical model introduced in the previous section can be used to investigate the image formation of the specific sample with known size and refractive index. To investigate the photothermal process, the temperature of the ``hot'' and ``cold'' states need to be solved. With a known size and refractive index as well as thermo-optic (dn/dT) and thermal-expansion (dr/dT) coefficient, the transient temperature profile for a particle placed on a silicon substrate can be simulated in COMSOL Multiphysics~\cite{Zhongming2017}. We perform the simulations in two steps (see details in supplementary 1 section 4). First, we numerically evaluate the absorbed mid-infrared power $ P_{abs} $ by a 500~nm PMMA particle. The total absorbed power is related to the mid-infrared beam intensity $I$ and the absorption cross-section $\sigma_{abs}$, $P_{abs} = \sigma_{abs}\cdot I$. Using the particle's optical parameters including the size and refractive index, the absorption cross-section is calculated. The mid-infrared beam intensity at the center of the IR focus is input from the experimentally measured power and beam size. The experimental details are explained in the supplementary information. In the second step, we calculate the transient temperature rise using the COMSOL's Heat Transfer in Solids module which takes the pre-calculated absorbed power as an input from the initial step. To do so, we define the geometry in which the bead is placed on top of the substrate. The bead is treated as a uniform heat source, which is reasonable as a result of the roughly uniform absorbed power distribution from the simulation result in the first step. The thermal diffusion process is calculated as the following equations: 

\begin{equation}
    \rho C_p \frac{\partial T}{\partial t} + \nabla \cdot \mathbf{q} = Q
\end{equation}

\begin{equation}
    \mathbf{q} = -k \nabla T 
\end{equation}

\noindent where $\rho$ is the density of the material, $C_p$ is the heat capacity at constant pressure, $T$ is temperature, $t$ is time, $k$ is the thermal conductivity. The COMSOL simulations can numerically solve these equations and obtain the temperature distribution in the time and space domain of the full system. 

\subsection{\sectionsize Sample preparation}
A 4" double-side polished silicon wafer with 500 µm thickness (University Wafer) is diced to 10 mm $\times$ 20 mm pieces. 500 nm PMMA beads (MMA500, Phosphorex) were diluted 10 times with deionized (DI) water and then spin-coated on the silicon substrate. The bacterial strains, \textit{S. aureus} ATCC 6538 and \textit{E. coli} BW 25113, used in this study were obtained from the Biodefense and Emerging Infections Research Resources Repository (BEI Resources) and the American Type Culture Collection (ATCC). To prepare bacterial samples for MIP imaging, bacterial strains were first cultured in cation-adjusted Mueller-Hinton Broth (MHB) (Thermo Fisher Scientific) media to reach the logarithmic phase. 1 mL of bacteria sample was centrifuged, washed twice with purified water, and then fixed by 10\% (w/v) formalin solution (Thermo Fisher Scientific). After centrifuging and washing with the purified water, 2 µL bacteria solution was deposited on a silicon substrate and dried at room temperature.

\begin{suppinfo}
\subsection{\sectionsize Supporting Information}
The Supporting Information is available free of charge on the \href{https://pubs.acs.org}{ACS Publications website}.

Bright-field and dark-field illumination photothermal signal details; Temperature dependence of photothermal signal; Backside infrared illumination optimization; COMSOL simulation details; Image acquisition and processing details. Figures S1-S17

\end{suppinfo}

\begin{acknowledgement}
This research was supported by National Institutes of Health R01GM126049, R35GM136223, R42CA224844, and R44EB027018 to J.X.C. C.Y and M.S.U acknowledge European Union’s Horizon 2020 Future and Emerging Technologies (No. 766466).
\end{acknowledgement}

\bibliography{references}

\end{document}